# An upper limit on the lateral vacancy diffusion length in diamond

J.O. Orwa[1*], K. Ganesan[2], J. Newnham[2], C. Santori[3], P. Barclay[3,4], K.M.C. Fu[3,5], R.G. Beausoleil[3], I. Aharonovich[6], B.A. Fairchild[2], P. Olivero[7], A.D. Greentree[2] and S. Prawer[2].

[1]*Department of Physics, Latrobe University, Bundoora, Victoria, Australia*

[2]*School of Physics, University of Melbourne, Parkville, Victoria 3010, Australia*

[3]*HP Laboratories, Palo Alto, California, USA*

[4]*Currently at Institute for Quantum Information Science, University of Calgary, Calgary, AB, T2N 1N3, Canada and NRC National Institute for Nanotechnology, 11421 Saskatchewan Drive NW, Edmonton, AB T6G 2M9, Canada*

[5]*Currently at Dept. of Physics, Dept. of Electrical Engineering, University of Washington, Seattle, WA 98195*

[6]*School of Engineering and Applied Sciences, Harvard University, Cambridge, MA, 02138, USA*

[7]*Experimental Physics Department and "Nanostructured Interfaces and Surfaces" Centre of Excellence, University of Torino, via P. Giuria 1, 10125 Torino, Italy and INFN Sezione di Torino, via P. Giuria 1, 10125 Torino, Italy*

**ABSTRACT**

Ion implantation is widely used to modify the structural, electrical and optical properties of materials. By appropriate masking, this technique can be used to define nano- and micro-structures. However, depending on the type of mask used, experiments have shown that vacancy-related substrate modification can be inferred tens of microns away from the edge of the mask used to define the implanted region. This could be due to fast diffusion of vacancies from the implanted area during annealing or to a geometric effect related to ion scattering around the mask edges. For quantum and single-atom devices, stray ion damage can be deleterious and must be minimized. In order to profile the distribution of implantation-induced damage, we have used the nitrogen-vacancy colour centre as a sensitive marker for vacancy concentration and distribution following MeV He ion implantation into diamond and annealing. Results show that helium atoms implanted through a mask clamped to the diamond surface are scattered underneath the mask to distances in the range of tens of micrometers from the mask edge. Implantation through a lithographically defined and deposited mask, with no spacing between the mask and the substrate, significantly reduces the scattering to ≤ 5 μm but does not eliminate it. These scattering distances are much larger than the theoretically estimated vacancy diffusion distance of ~260 nm under similar conditions. This paper shows that diffusion, upon annealing, of vacancies created by ion implantation in diamond is limited and the appearance of vacancies many tens of micrometers from the edge of the mask is due to scattering effects.

---

[*] Author to whom correspondence should be addressed. Electronic mail: J.Orwa@latrobe.edu.au.

1. **Introduction**

Ion implantation into a solid matrix is an important technology that can be used to modify the structural, optical and electronic properties of materials [1]. Where precise positioning of the ions is required, such as in high precision semiconductor and single atom device fabrication [2, 3], it is common to perform the implantation through a mask. Interaction of energetic ions with crystalline solids often results in the creation of vacancies, which can exist in isolation, as aggregates of two or more vacancies or as complexes with atoms. Such complexes can change the local potential landscape, leading to non-uniform device performance. For these reasons it is important to quantify the damage profiles obtained from ion implantation even down to the single vacancy level, especially for nanoscale quantum devices. In particular, assessing the distribution of vacancies with respect to the irradiated regions (mask edge) during the implantation process and after thermal annealing is of paramount importance. The presence of vacancies away from the implanted region (mask edge) could potentially arise from one or both of two processes. First, vacancies created within the implanted region could possibly diffuse out of the implanted region upon annealing, or alternatively, impinging ions scattered from the mask edges could result in the presence of vacancies far away from the implanted edge. We find here that the second scenario becomes more prominent as the gap between the mask and the substrate increases.

A number of studies have reported on the diffusion of atomic species as well as vacancies in diamond. For example, boron, oxygen, nitrogen and lithium have been reported to diffuse over distances of the order of 500 nm at temperatures of 860 °C [4]. Studies on diffusion of vacancies in diamond have provided conflicting estimates of the diffusion lengths, ranging from a few hundred nanometers in the vertical direction [5] to tens of micrometers in the lateral direction [6, 7]. In particular, ref. 12 reports lateral diffusion of vacancies during irradiation with electrons to distances exceeding 60 μm from the implantation edge.

The current work aims to shed light on this contradiction by distinguishing between vacancy distributions caused by diffusion upon annealing and those caused directly by ion scatter from the mask edges. To decouple the two processes, ion implantation needs to be carried out within a region with a well-defined spatial extent, separated from the rest of the material by a sharp boundary. This is best achieved by ion implantation through a mask, since the employment of conventional micro-beam ion scanning is significantly limited by the beam spot resolution. The simplest masking technique consists of a clamped mask such as cleaved silicon wafer or glass slide with a sharp edge. Alternatively, a lithographically defined and deposited (electroformed) mask can be used, although this technique is more expensive and requires more processing steps. While in the case of a clamped mask a gap is likely to exist between the mask and the substrate, no gaps should exist when the mask is electroformed. Consequently, different scattering effects are expected from the mask edge for implantations using the two types of masks, with the distribution range of scattered ions depending on the spacing between the mask and the substrate. To date, however, no special attention has been given to understanding the effects of scattering from the mask edges during ion implantation, leading to scattering processes sometimes being confused with diffusion.

During post-implantation annealing, a native nitrogen atom getters a vacancy to form a single NV center, which consists of a nitrogen atom close to a vacancy in a diamond lattice [8-13]. Vacancy distribution can thus be detected by monitoring the luminescent $NV^-$ distribution. The experimental results from the two types of masks are compared with numerical simulations for interaction of implanted ions with the mask edges.

## 2. Experimental

Three Sumitomo type Ib diamond samples grown by high pressure high temperature (HPHT) technique were implanted with 2 MeV He ions directed at an angle of 3º with respect to the mask edge (see Fig. 4(a)) in order to avoid channeling. The projected range of the ions was $3.51 \pm 0.12$ μm, according to SRIM Monte Carlo simulations [14]. Sample 1 was implanted to a fluence of $1\times10^{17}$ cm$^{-2}$ using a clamped sharp edged cleaved Si wafer as a mask. Sample 2 was implanted to a fluence of $1\times10^{17}$ cm$^{-2}$ through a clamped TEM grid while sample 3 was implanted in two different regions to fluences of $1\times10^{16}$ cm$^{-2}$ and $1\times10^{17}$ cm$^{-2}$, respectively, through a 5 μm thick electroformed gold mask (see Fig. 1). Fig. 1(a) is an SEM image showing the electroformed gold mask around the edges of a ~100×100 μm$^2$ exposed region of the diamond sample, while Fig. 1(b) is a depth profile along the dotted arrow in Fig. 1(a) showing that the thickness of the gold mask was slightly larger than 5 μm. Since the thickness of the gold mask is well above the projected range of 2.88 μm for 2 MeV He in gold, as determined by SRIM, the diamond sample was only exposed to the implanted He in the unmasked regions. This is confirmed in Fig. 1(c), which is an optical image of the sample taken after implantation with He to a fluence of $1\times10^{17}$ cm$^{-2}$ showing a darkened implanted region at the centre separated from the unimplanted surrounding region by a sharp edge. Fig. 1(d) is a height profile of the implanted and annealed (see annealing conditions for sample 3 in the next paragraph) surface showing an implantation-induced swelling of ~130 nm, which agrees with previous reports [6, 15, 16]. Sample 1 and sample 3 were annealed in forming gas (4% H$_2$ in Ar) at 900 ºC for 1 h followed by photo-luminescence (PL) analysis across the implanted edge to determine the NV$^-$ distribution. Prior to annealing, half of the irradiated region of sample 2 was etched off to a depth of ~1.4 µm using $O_2$/Ar plasma. This was followed by imaging and PL mapping to determine the pre-annealing distribution of NV⁻ or GR1 in the etched and unetched regions. The sample was then annealed at 900 °C for 1 h in forming gas before re-imaging to determine the post-annealing distribution of the NV⁻ and GR1 in the two regions.

The PL spectra were taken in steps of 0.5 µm along a line across the implanted region represented by the dotted arrow in Fig. 1(c). The intensities of the PL spectra were determined by integrating the area under the zero phonon line of the NV⁻ at 637 nm. These were then plotted to give an indication of the NV⁻ distribution and, hence, the vacancy diffusion.

## 3. Theoretical Estimates

An estimate of the expected diffusion distance can be obtained from the equation

$$D = D_0 \exp\left(\frac{-E_a}{kT}\right), \tag{1}$$

where $D$ is the diffusion coefficient, $E_a$ is the activation energy for vacancy diffusion, $k$ is the Boltzmann constant, $T$ is the temperature in K and $D_0$ is a pre-factor, usually obtained from an Arrhenius plot of experimental data. As an estimate, we use the recently reported value of $D_0 = 3.6 \times 10^{-6}$ cm² s⁻¹ for vacancy diffusion near a diamond surface [17], bearing in mind that the rate of vacancy diffusion near a diamond surface is higher than in the bulk. Several authors have previously reported values ranging from 1.7-4 eV [18-21] for the activation energy for diffusion of vacancies in bulk diamond. Using the minimum reported value for the activation energy (1.7 eV) and an annealing temperature of 900 °C, equation (1) results in a diffusion coefficient of $D = 1.82\times10^{-13}$ cm$^2$ s$^{-1}$. The diffusion distance, $d$, can be obtained from the equation:

$$d \approx \sqrt{Dt}, \tag{2}$$

where $t$ is the annealing time in seconds. Using the above value for the diffusion coefficient and the annealing time of 3600 s, equation (2) gives a diffusion distance of ~260 nm. Since we have used a pre-factor for vacancy diffusion near the surface, which is larger than for bulk, and the lowest reported diffusion coefficient, the diffusion distance obtained above is an overestimate and represents an upper limit for vacancy diffusion in bulk diamond upon annealing at 900 °C.

## 4. Results and Analysis

Fig. 2 shows NV$^-$ PL intensity as a function of position near the edge of implanted region for samples 1 and 3. The curves labeled "1e16-L" and "1e17-L" were obtained from samples implanted to fluences of $1\times10^{16}$ cm$^{-2}$ and $1\times10^{17}$ cm$^{-2}$, respectively, using electroformed gold masks (sample 3) while the curve labeled "1e17-P" was obtained from a sample implanted to a fluence of $1\times10^{17}$ cm$^{-2}$ (sample 1) through a cleaved Si wafer mask clamped to the diamond surface. The NV$^-$ intensity distribution for sample 1, following 1 h anneal, shows a rise in the NV$^-$ intensity just outside the edge of the implanted region followed by a gradual drop, extending to a distance $d_P = 64$ µm away from the implanted edge, where it reduces to the background intensity level. As shown in Fig. 2, when an electroformed mask is used (sample 3), the NV$^-$ intensity drops rapidly to the background level at a distance $d_L$, which is approximately less than or equal to 5 μm from the edge of the mask. Since the annealing time and annealing environments are the same, the vacancy diffusion rates should also be the same for the two samples. Thus the large discrepancy in the NV$^-$ distribution between the two samples suggests the presence of additional mechanisms which depend on the type of mask used. We suggest and show later that one such additional mechanism involves scattering of implanted He ions at the mask edge to distances underneath the mask that depend on the gap between the mask and the substrate and the angle the beam makes with the mask wall. Except in an ideal system in which the ion beam is parallel to the mask wall, there will always be some ions implanting through a small section of the mask near the sample surface. The scattered He ions directly create vacancies at their locations of impact, which explains the presence of NV$^-$ centres tens of micrometers away from the edge of the clamped mask (sample 1). This explanation also accounts for the observed limited distribution of the NV$^-$ from the mask edge for sample 3, where no gap is expected between the mask and the substrate.

Annealing of sample 3 for a further 4 h at 900 $^o$C in forming gas, then at 1100 $^o$C for 1 h in forming gas and finally at 1400 $^o$C for 1 h in vacuum did not show any indication of further redistribution of vacancies. We have suggested above that the long tail of NV$^-$ luminescence observed in Fig. 2 for sample 1 is due to scattering of implanted ions at the mask edge and not due to vacancy diffusion from the implanted region. This is further supported by results from an experiment in which we etched half of the implanted region of sample 2 to a depth of 1.4 μm prior to annealing. Figs. 3(a-b, d-e), are PL confocal maps obtained by using 532 nm laser excitation and broadband collection in the range of ~650-800 nm. Note that the dark regions in (a) and (b) represent the regions that were masked by the TEM grid. The spectra in Figs. 3(c) and 3(f) were obtained from the correspondingly labeled regions in the intensity maps of the unannealed and annealed samples, respectively. Fig 3(c) clearly shows that before annealing, PL related to the vacancy defect (GR1) was present in unimplanted areas even before the sample was annealed. This result strongly suggests that scattering occurs at the mask edges, as confirmed by the absence of the GR1 peak at a similar position in the etched region (see spectrum labeled "d" in Fig 3(c)). Such vacancy tails were also absent from a pristine region of the sample far away from the implanted region (see spectrum labeled "e" in Fig. 3(c)). Fig. 3(d) further shows that, after annealing, both $NV^-$ and $NV^0$ tails were present in unimplanted areas of the unetched region (see spectrum labeled "k" in Fig. 3(f)), but were absent in corresponding areas in the etched portion of the sample (see spectrum labeled "l" in Fig. 3(f)).

Results of the etching experiment, combined with the GR1 peak detected in an unimplanted region close to the surface before annealing, clearly indicate that ion scattering occurs during implantation. It further suggests that the NV tails observed in the unimplanted areas are due to vacancies that were created by scattering of incident He ions at the edges of the mask. Scattering to distances tens of micrometers away from the mask edge can occur only if a finite gap exists between the mask and the diamond substrate. Although the etch depth was less than the ion range in the directly implanted region, it was still sufficient to remove the $NV^-$ tails. This indicates that the ions creating the tails either have less total energy or are traveling at a steep angle from vertical. Ions scattered from the edges of the mask during implantation would have lower energy as well as travel at steep angles, which is consistent with the observed results.

5. **Discussion**

It was shown in Fig. 2 that the use of an electroformed mask significantly reduces the distribution of vacancies from the mask edge ($\leq 5$ μm) compared to a clamped mask (64 μm). However, the diffusion distance obtained when the electroformed mask is used is still about an order of magnitude larger than the theoretically estimated maximum diffusion distance of ~260 nm based on an activation energy of 1.7 eV. We conclude that, even for the electroformed mask where no gaps exist between the mask and the substrate, mechanisms other than diffusion are still present and account for the observed distribution of vacancies. In trying to find an explanation for the finite vacancy distribution even when no gap exists, we note that lateral straggling for 2 MeV He ions is only 113 nm and therefore cannot account for this effect. The laser spot size used in the PL measurements was 1 μm and measurements were taken at steps of 0.5 μm. Thus the spatial resolution in PL mapping is less than 1 μm and therefore, once again, cannot account for the distribution range observed. A third and most likely possibility is that some scattering still occurs at the edge of the mask, irrespective of the absence of any gaps.

To ascertain that even the short tails of $NV^-$ distribution in sample 3 (see Fig. 2) were due to vacancies created by scattered He ions during implantation, we used Geant4 (version 4.9.3) [22, 23] to simulate scattering of 2 MeV He ions directed at an angle of 3º into the edge of a clamped Si wafer mask as shown in Fig. 4 (a). The mask-target distance was varied during the simulation to spacings of 0, 10, 20, 50 and 100 μm, where a spacing of zero is tantamount to using an electroformed mask. He ions clip the edge of the clamped silicon wafer mask and are stopped or scattered. Ions that do not hit the mask are not simulated. Scattered ions hit the diamond target and come to rest with a distribution that depends on the mask-target spacing as shown in Fig. 4(b).

Even for an electroformed mask where the gap between the mask and substrate is zero, Fig. 2 shows that He ions still scatter underneath the mask to a distance of 2.4 µm, which is comparable to the NV$^-$ profile seen in Fig. 2 for sample 3. Thus, while the choice of a mask is important in reducing the amount of ion scatter at the mask edge, some scattering still remains even in the absence of a gap, making it difficult to ascertain the contribution of vacancy diffusion to the observed NV$^-$ distribution.

## 6. Summary and conclusions

We have used the luminescence from NV$^-$ centers to compare the distribution of vacancies in diamond caused by implantation of He atoms through a clamped mask and an electroformed mask. The results show that the employment of a clamped mask results in significant scattering of ions into the masked region of the substrate during implantation, causing the creation of vacancies at distances of tens of micrometers from the implanted edge, depending on the size of the gap. Using an electroformed mask, significantly reduces the scattering range to ≤ 5 µm, which is still substantially higher than the theoretically estimated upper limit for diffusion length of ~260 nm.

The results we have presented clearly show that the choice of a mask is an important parameter in determining the outcome of an ion implantation process. In particular we have shown that the unavoidable presence of a gap between a clamped mask and the substrate can lead to scattering of the implanted ions underneath the mask resulting in the creation of vacancies tens of micrometers away from the implanted edge, depending on the size of the gap. We have shown experimentally and with numerical simulations that such scattering can be significantly reduced by using electroformed masks instead. However, the scattering distance for electroformed masks is still larger than the estimated diffusion distance by more than an order of magnitude. Only an ideal system with parallel ion beam and mask wall will avoid ions implanting through a small section of the mask, near the sample surface, depending on the relative angle between them. We suggest that many previous results reporting long-distance vacancy diffusion in diamond may, in fact, have erroneously confused vacancies caused by scattering with diffusion.

**Acknowledgements**

This work was supported by the Australian Research Council (ARC) and the International Science Linkages Program of the Australian Department of Innovation, Industry, Science and Research (Project No. CG090191). A.D.G. acknowledges the Australian Research Council for financial support under Project No. DP0880466. A portion of this material is based upon work supported by the Defense Advanced Research Projects Agency under Award No. HR0011-09-1-0006 and The Regents of the University of California.

**References**

[1] M. Nastasi and J.W. Mayer, "Ion Implantation and Synthesis of Materials", Springer-Verlag, Berlin Heidelberg, 2006.

[2] Jessica A Van Donkelaar[1], Andrew D Greentree, Andrew D C Alves, Lenneke M Jong, Lloyd C L Hollenberg and David N Jamieson, New J.Phys. **12** (2010) 065016

[3] Alves A D C, Van Donkelaar J, Rubanov S, Reichart P and Jamieson D N 2008 Scanning transmission ion microscopy of nanoscale apertures *J. Korean Phys. Soc.* **53** 3704

[4] G. Popovici, R.G. Wilson, T. Sung, M.A. Prelas and S. Khasawinah, J. Appl. Phys **77** (1995) 5103.

[5] C. Santori, P.E. Barclay, K.M.C. Fu, R.G. Beausoleil, Phys. Rev. B **79** (2009) 125313.

[6] A. Gippius, R. Khmelnitskiy, V. Dravin, and S. Tkachenko, Diam. Relat. Mater. **8**, 1631 1999.

[7] J.W. Steeds, W. Sullivan, A. Wotherspoon and J.M. Hayes, J. Phys. Condens. Matter, **21** (2009) 364219.

[8] R.T. Harley, M.J. Henderson and R.M. Macfarlane, J. Phys. C: Solid State Phys. **17** (1984) L233.

[9] N.R.S. Reddy, N.B. Manson and E.R. Krausz, J. Lumin. **38** (1987) 46.

[10] A.D. Greentree, B.A. Fairchild, F.M. Hossain and S. Prawer, Mater. Today, **11** (2008) 22.

[11] Y. Mita, Phys. Rev. B, **53** (1996) 11360.

[12] A.Wotherspoon, J. Steeds, B. Catmull and J. Butler, Diamond Relat. Mater. **12** (2003) 652.

[13] G. Davies and M.F. Hamer, Proc. R. Soc. London Ser. A, **384** (1976) 285.

[14] http://www.ge.infn.it/~corvi/doc/software/srim2003/SRIM.htm

[15] F. Bosia, S. Calusi, L. Giuntini, S. Lagomarsino, A. Lo Giudice, M. Mass, P. Olivero, F. Picollo, S. Sciortino, A. Sordini, M. Vannoni and E. Vittone, Nucl. Instr. Meth. in Phys. Res. B **268** (2010) 299.

[16] A.V. Khomich, R. Khmelnitskiy, V. Dravin, A. Gippius, E. Zavedeev and I. Vlasov, Physics of the Solid State, **49** (2007) 1661.

[17] X.J. Hu, Y.B. Dai, R.B. Li, H.S. Shen and X.C. He, Sol. State Comm. 122 (2002) 45

[18] A Mainwood, Phys. Status Sol. (A) 172 (1999) 25

[19] S.J. Breur, P.R. Briddon, Phys. Rev. B 51 (1995) 6984.

[20] J. Bernhole, A. Antonelli, T.M. Del Sole, Phys. Rev. Lett. 61 (1988) 2689.

[21] J.O. Orwa, K.W. Nugent**,** D.N. Jamieson and S. Prawer, Phys. Rev. B, **62** (2000) 5461.

[22] S. Agostinelli, J. Allison, K. Amako, J. Apostolakis, H. Araujo, Arce, et. al., Nucl. Instr. Meth. in Phys. Res. A, **506 (**2003) 250 http://geant4.cern.ch/

[23] Mendenhall, M. & Weller, R., Nucl. Instr. Meth. in Phys. Res. B, **227** (2005) 420.

**List of Figures**

**Fig. 1** SEM image showing (a) - the electroformed gold mask; (b) – height profile across the mask along the dotted arrow shown in (a). The profile shows that the gold mask is a little over 5 μm thick, which is more than enough to stop 2 MeV He from reaching the diamond substrate underneath. (c) is an optical image showing the region implanted with He to a fluence of $1\times10^{17}$ ions/cm$^2$ after annealing at 900 °C for 1 h. (d) – height profile along the dotted arrow in (c) showing a swelling of 130 nm in the implanted region. The dotted arrow also represents a line along which NV$^-$ profile was recorded in order to provide a measure for vacancy diffusion (see Fig. 2). The exposed mask region in (a) and the implanted region in (c) are both squares. The rectangular appearance in (a) is due to a different camera angle.

**Fig. 2** NV$^-$ profile along a line across the implanted region for a sample implanted with 2 MeV He to a fluence of $1\times10^{17}$ ions/cm$^2$ through a clamped Si wafer mask (1e17-P) and for samples implanted with 2 MeV He through electroformed gold masks to fluences of $1\times10^{16}$ ions/cm$^2$ (1e16-L) and $1\times10^{17}$ ions/cm$^2$ (1e17-L). Both samples were annealed at 900 °C for 1 h. The long tail of NV$^-$ seen in the sample implanted through the cleaved Si mask is absent in the profile for the samples implanted through the electroformed gold mask. The dotted vertical line at ~95 μm represents the edges of the implanted regions while the dotted vertical lines at ~100 and 160 μm represent positions where the NV$^-$ intensity reduces to the background level for samples implanted through electroformed gold mask and clamped Si wafer mask, respectively. $d_L$ and $d_P$ are the corresponding distances from the edge of the implanted region. The plots have been displaced vertically for clarity.

**Fig. 3** Confocal maps and PL spectra of 2 MeV He implanted to a fluence of $1\times10^{17}$ ions/cm$^2$ through a TEM grid into a Sumitomo HPHT diamond showing the distribution of GR1, NV$^-$ and NV$^0$ both before and after annealing in the etched and unetched regions. The etching was carried out to a depth of 1.4 μm. The confocal maps and spectra were obtained at 10 K by using a 532 nm excitation laser, collecting through a broadband filter that covered from ~ 650 – 800 nm. (a) – Unetched region before annealing shows presence of vacancy (GR1) tails. (b) – etched region before annealing does not show presence of vacancy tails; (c) – PL spectra taken from the labeled points in the unetched and etched regions before annealing. The spectra have been individually scaled and vertically adjusted for clarity; (d) – unetched region after annealing shows presence of NV tails; (e) – etched region after annealing does not show presence of NV tails. In both the unetched and etched regions, there is a bright ring around the implanted region, ~1 μm wide; (f) – PL spectra taken from the labeled points in the unetched and etched regions after annealing. The spectra have been individually scaled and vertically adjusted for clarity. The peaks at 532 nm and 572 nm in (c) and (f) are the laser line and the diamond Raman line, respectively. The 20 μm vertical scale bar shown in (a) applies to all the confocal maps in Fig. 3.

**Fig. 4** (a) – Geant 4 simulation showing a thin 2 MeV He beam scattered off the edge of a clamped Si mask separated from the substrate by a 50 μm gap. He beam is seen scattered in all directions along the surface of the diamond substrate, including underneath the mask. The incident beam is aimed at an angle of 3° into the face of the mask. (b) – Dependence of the distribution of scattered ions on the gap between the mask and the diamond substrate. Larger gaps allow ions to travel further under the mask before hitting the target.

**Figures**

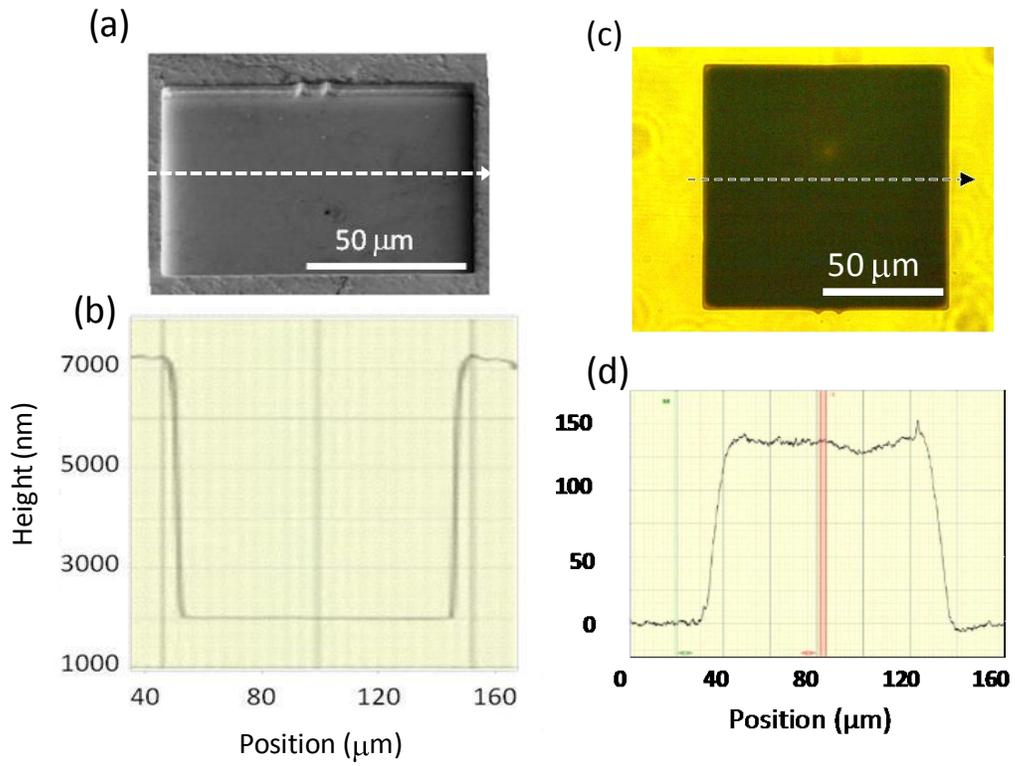

Fig. 1

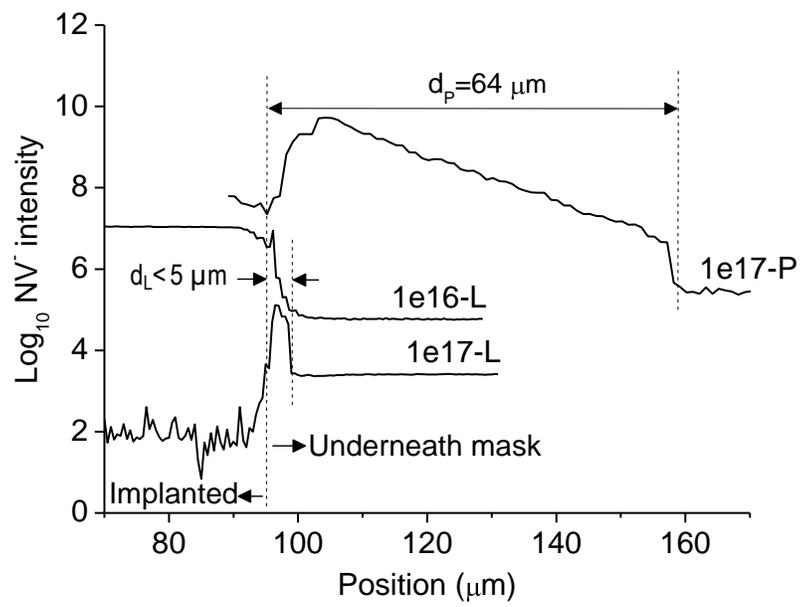

Fig. 2

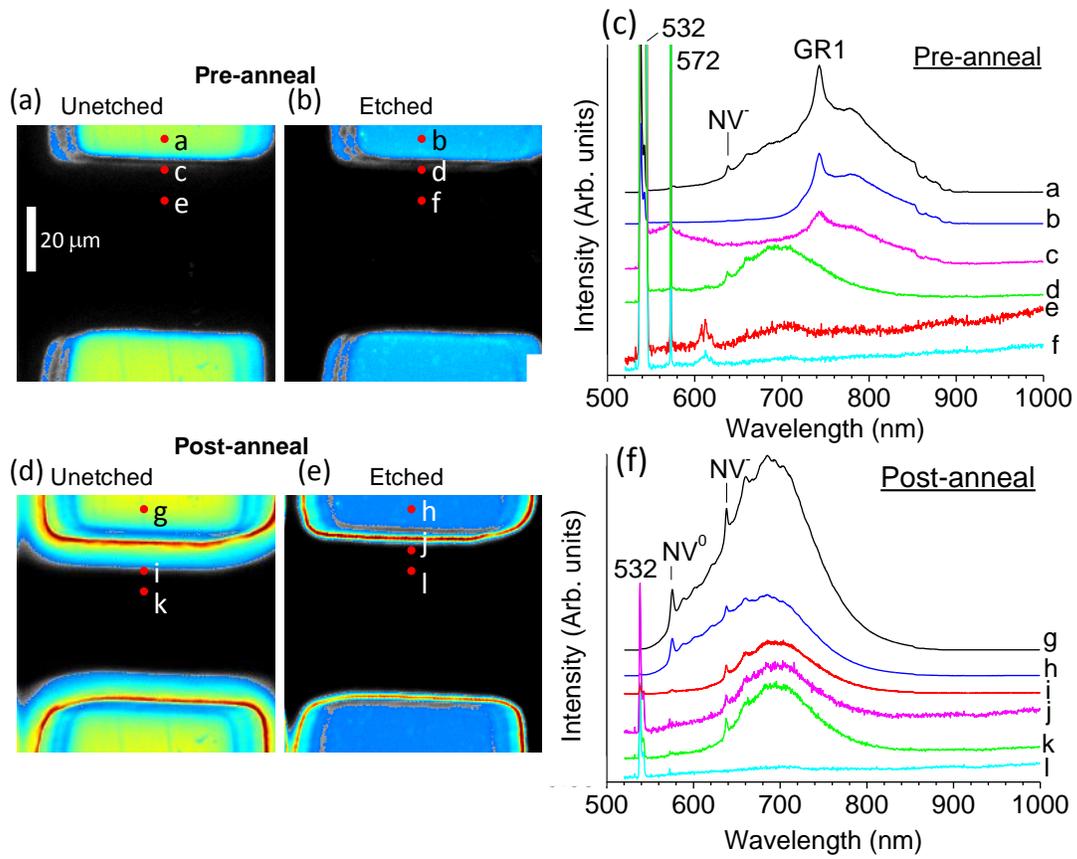

Fig. 3

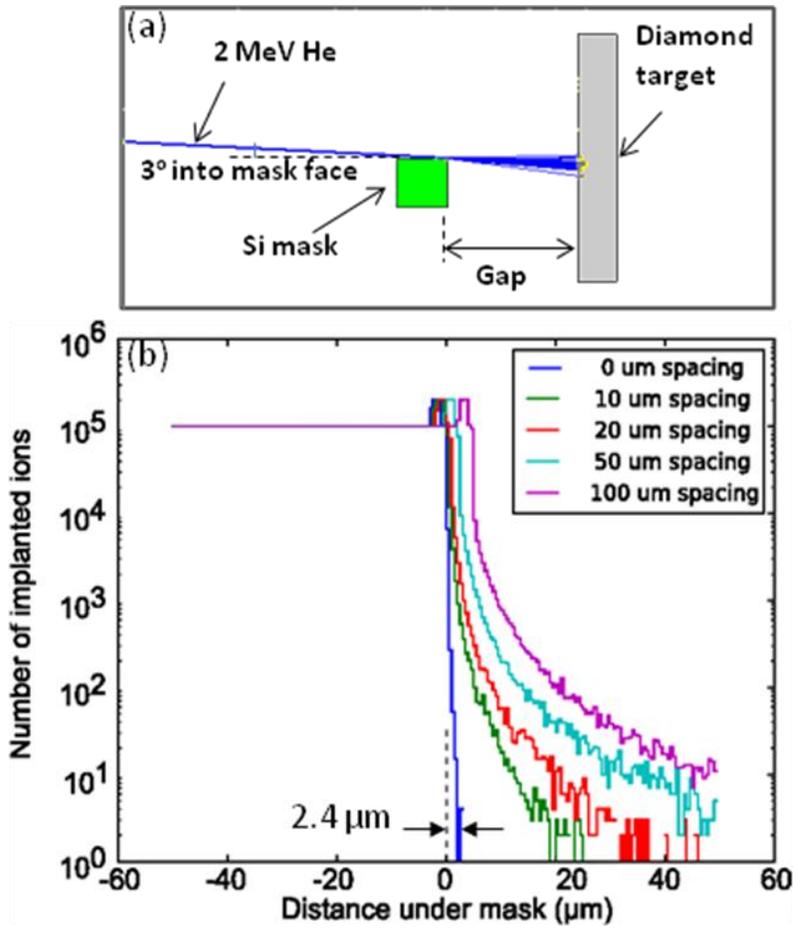

Fig. 4